\documentclass[english,12pt]{article}
\usepackage{array}
\usepackage{graphicx}
\usepackage{amssymb}
\usepackage[english]{babel}
\usepackage{amsmath}
\usepackage{multirow}
\usepackage{prettyref}
\usepackage{babel}
\usepackage{units}
\usepackage[latin1]{inputenc}
\usepackage{amsfonts}
\usepackage{amssymb}
\usepackage{babel}
\usepackage{color}
\usepackage{cite}
\def\@fmsl@sh#1#2#3{\m@th\ooalign{$\hfil#1\mkern#2/\hfil$\crcr$#1#3$}}
 \def\eq#1\en{\begin{equation}#1\end{equation}}
\def\s[#1,#2]{[#1\stackrel{\star}{,}#2]}
\def\sx[#1,#2]{[#1\stackrel{\star_{x}}{,}#2]}

\newcommand{\nc}{\newcommand}
\nc{\beq}{\begin{equation}}
\nc{\eeq}{\end{equation}}
\nc{\beqa}{\begin{eqnarray}}
\nc{\eeqa}{\end{eqnarray}}

\def\bc{\begin{center}}
\def\ec{\end{center}}

\def\gsim{\mathrel{\mathpalette\atversim>}}

\def\bc{\begin{center}}
\def\ec{\end{center}}

\def\gsim{\mathrel{\rlap{\lower4pt\hbox{\hskip1pt$\sim$}}

    \raise1pt\hbox{$>$}}}       

\def\gsim{\mathrel{\rlap{\lower4pt\hbox{\hskip1pt$\sim$}}
    \raise1pt\hbox{$>$}}}       

\begin{document}
\makeatletter
\def\fmslash{\@ifnextchar[{\fmsl@sh}{\fmsl@sh[0mu]}}
\def\fmsl@sh[#1]#2{%
  \mathchoice
    {\@fmsl@sh\displaystyle{#1}{#2}}%
    {\@fmsl@sh\textstyle{#1}{#2}}%
    {\@fmsl@sh\scriptstyle{#1}{#2}}%
    {\@fmsl@sh\scriptscriptstyle{#1}{#2}}}
\def\@fmsl@sh#1#2#3{\m@th\ooalign{$\hfil#1\mkern#2/\hfil$\crcr$#1#3$}}
\makeatother

\thispagestyle{empty}
\begin{titlepage}
\boldmath
\begin{center}
  \Large {\bf Quantum Gravitational Corrections to the Entropy of a Schwarzschild Black Hole}
    \end{center}
\unboldmath
\vspace{0.2cm}
\begin{center}
{{\large Xavier~Calmet}\footnote{E-mail: x.calmet@sussex.ac.uk},
{\large and}  {\large  Folkert~Kuipers}\footnote{E-mail: f.kuipers@sussex.ac.uk}}
 \end{center}
\begin{center}
{\sl Department of Physics and Astronomy,
University of Sussex, Brighton, BN1 9QH, United Kingdom
}\\
\end{center}
\vspace{5cm}
\begin{abstract}
\noindent
We calculate quantum gravitational corrections to the entropy of black holes using the Wald entropy formula within an effective field theory approach to quantum gravity. The corrections to the entropy are calculated to second order in curvature and we calculate a subset of those at third order. We show that, at third order in curvature, interesting issues appear that had not been considered previously in the literature. The fact that the Schwarzschild metric receives corrections at this order in the curvature expansion has important implications for the entropy calculation. Indeed, the horizon radius and the temperature receive corrections. These corrections need to be carefully considered when calculating the Wald entropy.
\end{abstract}
\vspace{5cm}
\end{titlepage}



\newpage

Black holes are fascinating objects for many different reasons. Hawking's groundbreaking intuition that black holes are not black but have a radiation spectrum that is very similar to that of a black body makes black holes an ideal laboratory to investigate the interplay between quantum mechanics, gravity and thermodynamics. This has led to the notion of Bekenstein-Hawking entropy or black hole entropy which has attracted much attention over the last almost 50 years. The calculation of quantum corrections to this entropy has been the subject of many publications, see e.g. \cite{Solodukhin:2011gn,Wall:2018ydq} for  reviews.

In this work we revisit the calculation of the entropy of a Schwarzschild black hole in quantum gravity and identify new important subtleties that have been overlooked in previous calculations. To be very specific, we use effective field theoretical methods to calculate quantum gravitational corrections to the entropy of this black hole using the Wald entropy formula \cite{Wald:1993nt}. We highlight new intriguing relations between the quantum corrections to the entropy, the Euler characteristic and quantum corrections to the metric of the Schwarzschild black hole. Previous calculations within the effective theory approach to quantum gravity  \cite{Fursaev:1994te,El-Menoufi:2015cqw,El-Menoufi:2017kew} have used the Euclidean path integral formulation of the entropy. We present a systematic approach that can easily be extended to any order in perturbation theory or to any black hole metric.

The Wald approach to the calculation of a black hole entropy is very elegant and does not involve the Wick rotation to Euclidean time which is known to be tricky in quantum gravity. The Wald entropy formula reads \cite{Wald:1993nt}
\begin{eqnarray}
S_{Wald} =  -2\pi \int d\Sigma \  \epsilon_{\mu\nu} \epsilon_{\rho\sigma}
	 \frac{\partial \mathcal{L}}{\partial R_{\mu\nu\rho\sigma}}\Big|_{r=r_H} ,
\end{eqnarray}
where $d\Sigma=r^2 \sin \theta d\theta d\phi$, $L$ is the Lagrangian of the model, $R^{\mu\nu\rho\sigma}$ is the Riemann tensor and $r_H$ is the horizon radius. Furthermore, $\epsilon_{\mu\nu}\epsilon^{\mu\nu}=-2$, $\epsilon_{\mu\nu}=-\epsilon_{\nu\mu}$. The integral is over the perimeter of the horizon of the black hole and we thus need to determine the location of the horizon with radius $r_H$. This is our first observation: to calculate the entropy of the black hole, we do not only need the Lagrangian of the gravitational action, but we also need to verify whether the metric receives quantum corrections as these could impact the position of the horizon. This important point had simply been overlooked in previous calculations for Schwarzschild black holes.

As explained before, we are using the effective action to quantum gravity~\cite{Weinberg:1980gg, Barvinsky:1984jd,Barvinsky:1985an,Barvinsky:1987uw,Barvinsky:1990up,Buchbinder:1992rb,Donoghue:1994dn}. At second order in curvature, one has
\begin{align}\label{EFTaction}
S_{\text{EFT}} = \int \sqrt{|g|}d^4x  \left( \frac{R}{16\pi G_N} + c_1(\mu) R^2 + c_2(\mu) R_{\mu\nu} R^{\mu\nu} + c_3(\mu) R_{\mu\nu\rho\sigma} R^{\mu\nu\rho\sigma} + \mathcal{L}_m \right) \ ,
\end{align}
for the local part of the action and the nonlocal part is given by
\begin{align}\label{nonlocalaction}
	\Gamma_{\text{NL}}^{\scriptstyle{(2)}}  = - \int  \sqrt{|g|}d^4x \left[ \alpha R \ln\left(\frac{\Box}{\mu^2}\right)R + \beta R_{\mu\nu} \ln\left(\frac{\Box}{\mu^2}\right) R^{\mu\nu} + \gamma R_{\mu\nu\alpha\beta} \ln\left(\frac{\Box}{\mu^2}\right)R^{\mu\nu\alpha\beta} \right],
	\end{align}
	where $\Box := g^{\mu\nu} \nabla_\mu \nabla_\nu$.

It is straightforward to show \cite{Calmet:2017qqa,Calmet:2018elv} that there are no corrections up to second order in curvature to the Schwarzschild metric using the non-local Gauss-Bonnet identity \cite{Calmet:2018elv}
\begin{eqnarray}
\int   \sqrt{|g|} d^4x R_{\mu\nu\alpha\beta}\left (c_3(\mu)-\gamma  \ln\left(\frac{\Box}{\mu^2}\right) \right) R^{\mu\nu\alpha\beta} &=&
+4 \int  \sqrt{|g|} d^4xR_{\mu\nu}\left (c_3(\mu)-\gamma  \ln\left(\frac{\Box}{\mu^2}\right)\right)R^{\mu\nu}  \nonumber \\ &&
-\int   \sqrt{|g|} d^4xR\left (c_3(\mu)-\gamma  \ln\left(\frac{\Box}{\mu^2}\right)\right)R \nonumber \\ && + {\cal O}(R^3)+ {\rm  boundary \ terms.}
\end{eqnarray}
This identity can be proven using
 \cite{Knorr:2019atm,Donoghue:2015nba}
\begin{eqnarray}
\log \frac{\Box}{\mu^2}=\int_0^\infty ds \frac{e^{-s}-e^{-s\frac{\Box}{\mu^2}} }{s}
\end{eqnarray}
and  \cite{Barvinsky:1990up}
\begin{eqnarray}
\Box R^{\alpha\beta\mu\nu}&=&\nabla^\mu\nabla^\alpha R^{\nu\beta} -\nabla^\nu\nabla^\alpha R^{\mu\beta}-\nabla^\mu\nabla^\beta R^{\nu\alpha}+
+\nabla^\nu\nabla^\beta R^{\mu\alpha}\nonumber \\ && -4  R^{\alpha \ [\mu}_{\ \sigma \ \lambda} R^{\beta\sigma\nu]\lambda}
+ 2 R^{ [ \mu}_{\ \lambda} R^{\alpha\beta\lambda\nu] }- R^{\alpha\beta}_{\ \ \sigma \lambda}R^{\mu\nu\sigma\lambda},
 \end{eqnarray}
which follows from the Bianchi identity. One obtains \cite{Knorr:2019atm,Deser:1986xr,Asorey:1996hz,Teixeira:2020kew}
\begin{eqnarray}
R_{\alpha\beta\mu\nu}\Box R^{\alpha\beta\mu\nu}= 4 R_{\alpha\beta\mu\nu}\nabla^\alpha \nabla^\mu R^{\beta\nu}+ {\cal O}(R^3).
 \end{eqnarray}
It is straightforward to generalize this result to higher power of the Laplacian. Inserting this relation into the Lagrangian and using partial integrations and the contracted Bianchi identity, we obtain the non-local Gauss-Bonnet identity. As the Riemann tensor can be eliminated from the dynamical part of the action at second order in curvature, we find that there are no corrections to the field equations at this order for vacuum solutions of general relativity \cite{Calmet:2018elv}.

As there are no corrections to the metric, the horizon radius is unchanged and we can calculate the Wald entropy at second order in a straightforward manner using (\ref{EFTaction}) and (\ref{nonlocalaction})\footnote{Note that we need to use this basis  for the calculation of the entropy, as we have not calculated the boundary term generated by Gauss-Bonnet identity explicitly.}
\begin{eqnarray} \label{2ndentropy}
S_{Wald}^{(2)}  &=&\frac{A}{4 G_N} +  64 \pi^2 c_3(\mu) +64 \pi^2 \gamma \left( \log \left (4 G^2_N M^2 \mu^2\right) -2 +2 \gamma_E \right)
\end{eqnarray}
where $A=16 \pi (G_N M)^2$ is the area of the black hole. A similar answer was obtained using  the Euclidean path integral formulation. Note that the entropy is renormalization group invariant and finite. As there are no corrections to the metric, the temperature remains unchanged and the classical relation $TdS=dM$ receives a quantum correction. Indeed we find $T dS= (1+\gamma 16 \pi /(G_N M^2))dM$.

A possible interpretation of this result is that the nonlocal quantum effects generate a pressure for the black hole. The first law of thermodynamics is then given by
\begin{eqnarray}
TdS-PdV=\left(1+\gamma \frac{16 \pi}{G_N M^2}\right) dM=  dM+\gamma \frac{16 \pi}{G_N M^2} dM,
\end{eqnarray}
where $P$ is the pressure of the black hole. Its volume is given by $V=4/3 \pi r_H^3$, where $r_H= 2 G_N M$ is the horizon radius.
We can then identify $TdS=dM$ and $\gamma 16 \pi/(G_N M^2) dM= - P dV$  with $dV=32 \pi G_N^3 M^2 dM$. We thus obtain
\begin{eqnarray} \label{pressure}
P=-\gamma \frac{1}{2 G_N^4 M^4},
\end{eqnarray}
which can be negative as $\gamma$ is positive for spin 0, 1/2 and 2 fields or positive as $\gamma$ is negative for spin 1 fields. Indeed, one finds  $\gamma_0=2/(11520 \pi^2)$ \cite{Deser:1974cz},  $\gamma_{1/2}=7/(11520 \pi^2)$ \cite{Deser:1974cz}, $\gamma_{1}=-26/(11520 \pi^2)$ \cite{Deser:1974cz} and $\gamma_{2}=424/(11520 \pi^2)$ \cite{Barvinsky:1984jd}. We note that Dolan had discussed the possibility that black holes would have a pressure \cite{Dolan:2011xt} in the context of gravitational models with a cosmological constant. It is remarkable that quantum gravity leads to a pressure for Schwarzschild black holes. Note that this is the main difference with previous results  \cite{Fursaev:1994te,El-Menoufi:2015cqw,El-Menoufi:2017kew} who did not study quantum corrections to the metric. Because there is no dynamical correction to the metric at this order in curvature, the interpretation of the correction to the entropy as a pressure term is forced upon us.

At third order in curvature, we need to add the following operators to the effective action
\begin{eqnarray}
{\cal L}^{(3)}=c_6 G_N  R^{\mu\nu}_{\ \ \alpha\sigma} R^{\alpha\sigma}_{\ \ \delta\gamma} R^{\delta\gamma}_{\ \ \mu\nu}  \ ,
\end{eqnarray}
where $c_6$ is dimensionless. As pointed out by Goroff and Sagnotti \cite{Goroff:1985th}, there is only one invariant involving only Riemann tensors in vacuum, as  $R_{\alpha\beta\gamma\delta} R^{\alpha \ \gamma}_{\ \epsilon \ \zeta} R^{\beta\epsilon\delta\zeta}$ can be rewritten in terms of $R^{\mu\nu}_{\ \ \alpha\sigma} R^{\alpha\sigma}_{\ \ \delta\gamma} R^{\delta\gamma}_{\ \ \mu\nu}$ and terms involving the Ricci scalar or Ricci tensors which both vanish in vacuum. There is a corresponding non-local operator $R^{\mu\nu}_{\ \ \alpha\sigma} \log{\Box} R^{\alpha\sigma}_{\ \ \delta\gamma}R^{\delta\gamma}_{\ \ \mu\nu}$. While the Wilson coefficient is known in a specific gauge \cite{Goroff:1985th}, it is not known for the unique effective action and we will thus neglect  this term.

The dimension six local operator  leads to  a correction to the metric. We find
\begin{eqnarray}
	ds^2 = - f(r) dt^2 + \frac{1}{g(r)} dr^2 + r^2 d\Omega^2
\end{eqnarray}
with
\begin{eqnarray}
	d\Omega^2 &=& d\theta^2 + \sin(\theta)^2 d\phi^2, \\
	f(r) &=& 1 - \frac{2 G_N M}{r} +  640 \pi c_6 \frac{G_N^5 M^3}{r^7},\\
	g(r) &=& 1 - \frac{2G_N  M}{r} + 128 \pi c_6 \frac{G_N^4 M^2 }{r^6} \left(27 - 49 \frac{G_N M}{r} \right).
\end{eqnarray}

The corrections to the metric implies a shift of the horizon radius
\begin{eqnarray}
r_H= 2 G_N M \left (1 - c_6 \frac{5 \pi}{G_N^2 M^4 }\right).
\end{eqnarray}
Clearly for astrophysical black holes the correction to the classical Schwarzschild radius goes to zero very quickly but it can be an order one correction for quantum black holes with masses of the order of the Planck scale.

The $\epsilon_{\mu\nu}$ tensors also need to be redefined. We have
\begin{equation}
	\epsilon_{\mu\nu} = \begin{cases}
	\sqrt{f(r)/g(r)} \qquad &{\rm if} \quad (\mu,\nu)=(t,r),\\
	- \sqrt{f(r)/g(r)} \qquad &{\rm if} \quad (\mu,\nu)=(r,t),\\
	0 \qquad &{\rm otherwise}.
	\end{cases}
\end{equation}
One can easily verify that  $\epsilon_{\mu\nu}\epsilon^{\mu\nu}=-2$, $\epsilon_{\mu\nu}=-\epsilon_{\nu\mu}$, and $\epsilon_{\mu\nu}=0$, if $\mu,\nu\neq t,r$.

At third order in curvature, we thus obtain the following correction to the entropy:
\begin{eqnarray}
S_{Wald}^{(3)} &=&S_{Wald}^{(2)}
+128 \pi^3 c_6 \frac{G_N}{A_{tot}} \ , \  \ \ \
\end{eqnarray}
where we neglect third order non-local terms which would compensate for the scale dependence of $c_6$ \footnote{We can estimate the magnitude of the non-local correction of the entropy (albeit in the de Donder gauge, the actual calculation in the unique effective action would be much more involved) using the result in \cite{Goroff:1985th} for the two-loop divergences of Einstein gravity
$ \Gamma_\infty = \frac{209}{2880 (4 \pi)^4} \frac{1}{\epsilon} \int d^4x \sqrt{-g} R^{\mu\nu}_{\ \ \alpha\sigma}  R^{\alpha\sigma}_{\ \ \delta\gamma}R^{\delta\gamma}_{\ \ \mu\nu}$. This divergent term fixes the renormalization group equation for $c_6$ and thus the Wilson coefficient of the term $R^{\mu\nu}_{\ \ \alpha\sigma} \log{\Box} R^{\alpha\sigma}_{\ \ \delta\gamma}R^{\delta\gamma}_{\ \ \mu\nu}$. For the entropy to be renormalization group invariant at third order in curvature, the non-local correction to the entropy must go as  $\frac{209}{2880 (4 \pi)^4} G_N/A_{tot} \log (4 G_N^2 M^2 \mu^2)$. These corrections are thus very small in comparison to those obtained in eq. (\ref{2ndentropy}).}. Note that while the dimension six operator has been considered before \cite{Solodukhin:2019xwx}, our result differs from that paper as the metric corrections were not taken into account in that work.

With corrections to the metric that deviate from the Schwarzschild solution, one may wonder whether the Euler characteristic given by
\begin{equation}
	\chi
	= \frac{1}{32\pi^2} \int_{0}^{1/T} dt_E \int_{r_H}^{\infty} dr \int_{0}^{\pi} d\theta \int_{0}^{2\pi} d\phi
	\sqrt{|g|} \left(R^2 - 4 R_{\mu\nu} R^{\mu\nu} + R_{\mu\nu\rho\sigma} R^{\mu\nu\rho\sigma} \right)
	\end{equation}
remains 2 for black holes. It is however easy to see that this is the case, because there is also a correction to the temperature which is given by
\begin{equation}
	T
	= \frac{\sqrt{f'(r_H) g'(r_H)}}{4\pi}
	= \frac{1}{8\pi G_N M} \left[1 + 2\pi c_6 \left(\frac{1}{G_N^2 M^4}\right) \right].
\end{equation}
With this in mind, it is easy to verify that $\chi=2$ is fulfilled, which is required for our results to be consistent.
 One can also easily verify that the thermodynamic law $TdS=dM$ holds at order of ${\cal O}(c_6)$ with the modified temperature and entropy. The non-local correction to the action at third order in curvature would lead to a contribution to the pressure which is much smaller than the seconder order correction obtained in eq. (\ref{pressure}). A back of the envelop calculation shows that, as expected, the third order curvature nonlocal correction to the pressure is suppressed by a factor $(G_N M^2)^{-1}$ in comparison to the leading second order term that we have calculated.

Our work has interesting implications for quantum black holes. The temperature of black holes can be seen as an indicator of how quantum a black hole is. A black hole with a mass of the order of ten times the reduced Planck mass $\bar M_P$ would still be a very good approximation and have a temperature close to its classical value
\begin{equation}
	T_{QBH}	= \frac{1}{8\pi G_N \bar M_P} \left[1 + 128 \pi^3 c_6 \frac{\bar M_P^4}{M_{QBH}^4} \right].
\end{equation}
Assuming that $c_6$ is of order unity, we see that the classical temperature receives an order one correction from the third order curvature term in the action for $M_{QBH}=\bar M_P$, but these corrections are very tiny for quantum black holes with masses of the order of $M_{QBH}\sim 10 \bar M_P$. This justifies the geometrical cross-section adopted for quantum black holes in the framework of low scale quantum gravity at colliders \cite{Calmet:2008dg,Meade:2007sz,Giddings:2001bu,Hsu:2002bd}. The semi-classical approximation appears to be an excellent one. Describing quantum black holes with the classical Schwarzschild metric is clearly a good approximation as well as long as their masses are larger than ${\cal O}(10 \bar M_P)$.

In this work we have calculated quantum gravitational corrections to the entropy of black holes using the Wald entropy formula within an effective theory approach to quantum gravity at third order in curvature.
We first have revisited the calculation of the entropy of black holes at second order in curvature and have found that the quantum gravitational correction to the entropy can be interpreted as a pressure term in the first law of thermodynamics for black holes. This pressure can be positive or negative depending on the field content of the theory. Furthermore, we have shown that at third order in curvature, there are interesting issues that had not been considered previously in the literature. The fact that the Schwarzschild metric receives corrections at this order in the curvature expansion has important implications for the entropy calculation. Indeed, the horizon radius and the temperature receive corrections. These corrections need to be carefully considered when calculating the Wald entropy, knowing the corrections to the Lagrangian is not enough. The reason why previous entropy calculations at second order in curvature match our results is that there are no correction to the Schwarzschild metric at that order. We can actually justify this result with our approach. Finally, our results have interesting consequences for the lightest black holes of Planckian masses \cite{Calmet:2014gya,Calmet:2015pea} which are much more classical than naively expected.

{\it Acknowledgments:}
We would like to thank Yong Xiao for very useful discussions. The work of X.C. is supported in part  by the Science and Technology Facilities Council (grants numbers ST/T00102X/1, ST/T006048/1 and ST/S002227/1). The work of F.K. is supported by a doctoral studentship of the Science and Technology Facilities Council. 



\baselineskip=1.6pt

\begin{thebibliography}{10}




\bibitem{Solodukhin:2011gn}
S.~N.~Solodukhin,
Living Rev. Rel. \textbf{14} (2011), 8
doi:10.12942/lrr-2011-8
[arXiv:1104.3712 [hep-th]].

\bibitem{Wall:2018ydq}
A.~C.~Wall,
[arXiv:1804.10610 [gr-qc]].

\bibitem{Wald:1993nt}
R.~M.~Wald,
Phys. Rev. D \textbf{48} (1993) no.8, 3427-3431
doi:10.1103/PhysRevD.48.R3427
[arXiv:gr-qc/9307038 [gr-qc]].


\bibitem{Fursaev:1994te}
D.~V.~Fursaev,
Phys. Rev. D \textbf{51}, 5352-5355 (1995)
doi:10.1103/PhysRevD.51.R5352
[arXiv:hep-th/9412161 [hep-th]].

\bibitem{El-Menoufi:2017kew}
B.~K.~El-Menoufi,
JHEP \textbf{08} (2017), 068
doi:10.1007/JHEP08(2017)068
[arXiv:1703.10178 [gr-qc]].



\bibitem{El-Menoufi:2015cqw}
B.~K.~El-Menoufi,
JHEP \textbf{05} (2016), 035
doi:10.1007/JHEP05(2016)035
[arXiv:1511.08816 [hep-th]].


\bibitem{Weinberg:1980gg}
S.~Weinberg,
``Ultraviolet Divergences In Quantum Theories Of Gravitation,''
\textit{General Relativity: An Einstein Centenery Survey},
Cambridge, UK, 790 (1980).


\bibitem{Barvinsky:1984jd}
A.~O.~Barvinsky and G.~A.~Vilkovisky,
Phys.\ Lett.\  {\bf 131B}, 313 (1983).
doi:10.1016/0370-2693(83)90506-3


\bibitem{Barvinsky:1985an}
A.~O.~Barvinsky and G.~A.~Vilkovisky,
Phys.\ Rept.\  {\bf 119}, 1 (1985).
doi:10.1016/0370-1573(85)90148-6


\bibitem{Barvinsky:1987uw}
A.~O.~Barvinsky and G.~A.~Vilkovisky,
Nucl.\ Phys.\ B {\bf 282}, 163 (1987).
doi:10.1016/0550-3213(87)90681-X


\bibitem{Barvinsky:1990up}
A.~O.~Barvinsky and G.~A.~Vilkovisky,
Nucl.\ Phys.\ B {\bf 333}, 471 (1990).
doi:10.1016/0550-3213(90)90047-H


\bibitem{Buchbinder:1992rb}
I.~L.~Buchbinder, S.~D.~Odintsov and I.~L.~Shapiro,
``Effective action in quantum gravity,''
Bristol, UK: IOP, 413 (1992).


\bibitem{Donoghue:1994dn}
J.~F.~Donoghue,
Phys.\ Rev.\ D {\bf 50}, 3874 (1994).
doi:10.1103/PhysRevD.50.3874

\bibitem{Calmet:2017qqa}
X.~Calmet and B.~K.~El-Menoufi,
Eur. Phys. J. C \textbf{77} (2017) no.4, 243
doi:10.1140/epjc/s10052-017-4802-0
[arXiv:1704.00261 [hep-th]].

\bibitem{Calmet:2018elv}
X.~Calmet,
Phys. Lett. B \textbf{787} (2018), 36-38
doi:10.1016/j.physletb.2018.10.040
[arXiv:1810.09719 [hep-th]].


\bibitem{Knorr:2019atm}
B.~Knorr, C.~Ripken and F.~Saueressig,
Class. Quant. Grav. \textbf{36}, no.23, 234001 (2019)
doi:10.1088/1361-6382/ab4a53
[arXiv:1907.02903 [hep-th]].

\bibitem{Donoghue:2015nba}
J.~F.~Donoghue and B.~K.~El-Menoufi,
JHEP \textbf{10}, 044 (2015)
doi:10.1007/JHEP10(2015)044
[arXiv:1507.06321 [hep-th]].



\bibitem{Deser:1986xr}
S.~Deser and A.~N.~Redlich,
Phys. Lett. B \textbf{176}, 350 (1986)
[erratum: Phys. Lett. B \textbf{186}, 461 (1987)]
doi:10.1016/0370-2693(86)90177-2

\bibitem{Asorey:1996hz}
M.~Asorey, J.~L.~Lopez and I.~L.~Shapiro,
Int. J. Mod. Phys. A \textbf{12}, 5711-5734 (1997)
doi:10.1142/S0217751X97002991
[arXiv:hep-th/9610006 [hep-th]].

\bibitem{Teixeira:2020kew}
P.~d.~Teixeira, I.~L.~Shapiro and T.~G.~Ribeiro,
Grav. Cosmol. \textbf{26}, no.3, 185-199 (2020)
doi:10.1134/S0202289320030123
[arXiv:2003.04503 [hep-th]].

\bibitem{Deser:1974cz}
S.~Deser and P.~van Nieuwenhuizen,
Phys. Rev. D \textbf{10}, 401 (1974)
doi:10.1103/PhysRevD.10.401

\bibitem{Dolan:2011xt}
B.~P.~Dolan,
Class. Quant. Grav. \textbf{28}, 235017 (2011)
doi:10.1088/0264-9381/28/23/235017
[arXiv:1106.6260 [gr-qc]].

\bibitem{Goroff:1985th}
M.~H.~Goroff and A.~Sagnotti,
Nucl. Phys. B \textbf{266} (1986), 709-736
doi:10.1016/0550-3213(86)90193-8


\bibitem{Solodukhin:2019xwx}
S.~N.~Solodukhin,
Phys. Lett. B \textbf{802} (2020), 135235
doi:10.1016/j.physletb.2020.135235
[arXiv:1907.07916 [hep-th]].

\bibitem{Calmet:2008dg}
X.~Calmet, W.~Gong and S.~D.~H.~Hsu,
Phys. Lett. B \textbf{668} (2008), 20-23
doi:10.1016/j.physletb.2008.08.011
[arXiv:0806.4605 [hep-ph]].

\bibitem{Meade:2007sz}
P.~Meade and L.~Randall,
JHEP \textbf{05} (2008), 003
doi:10.1088/1126-6708/2008/05/003
[arXiv:0708.3017 [hep-ph]].

\bibitem{Giddings:2001bu}
S.~B.~Giddings and S.~D.~Thomas,
Phys. Rev. D \textbf{65} (2002), 056010
doi:10.1103/PhysRevD.65.056010
[arXiv:hep-ph/0106219 [hep-ph]].

\bibitem{Hsu:2002bd}
S.~D.~H.~Hsu,
Phys. Lett. B \textbf{555} (2003), 92-98
doi:10.1016/S0370-2693(03)00012-1
[arXiv:hep-ph/0203154 [hep-ph]].

\bibitem{Calmet:2014gya}
X.~Calmet,
Mod. Phys. Lett. A \textbf{29} (2014) no.38, 1450204
doi:10.1142/S0217732314502046
[arXiv:1410.2807 [hep-th]].


\bibitem{Calmet:2015pea}
X.~Calmet and R.~Casadio,
Eur. Phys. J. C \textbf{75} (2015) no.9, 445
doi:10.1140/epjc/s10052-015-3668-2
[arXiv:1509.02055 [hep-th]].




\end{thebibliography}
\end{document}